\documentclass[twocolumn,nofootinbib,jcp,preprintnumbers,amsmath,amssymb]{revtex4-1}
\usepackage{natbib,hyperref}
\usepackage{amsmath}
\usepackage{graphicx}
\usepackage{dcolumn}
\usepackage{tabularx}
\usepackage{bm}
\usepackage{epsfig,color,xspace,multirow,xr,bbold}
\usepackage[all]{xy}
\usepackage{setspace}
\usepackage{url}
\usepackage{threeparttable, tablefootnote}
\newcommand{\RRef}[1]{Ref.~\onlinecite{#1}}

\usepackage{natbib,hyperref}
\usepackage{float}
\usepackage{placeins}
\usepackage[utf8]{inputenc}

\usepackage[none]{hyphenat}
\usepackage{calligra}

\begin{document}

\title[]{
Machine Learning Modeling of Materials with a Group-Subgroup Structure
}

\author{Prakriti Kayastha}
\author{Raghunathan Ramakrishnan}%
 \email{ramakrishnan@tifrh.res.in}
\affiliation{ 
Tata Institute of Fundamental Research, Centre for Interdisciplinary Sciences, Hyderabad 500107, India}

\date{\today}

\begin{abstract}
 Crystal structures connected by continuous phase transitions are linked through mathematical relations between crystallographic groups and their subgroups. In the present study, we introduce {\it group-subgroup machine learning} (GS-ML) and show that including materials with small unit cells in the training set decreases out-of-sample prediction errors for materials with large unit cells. GS-ML incurs the least training cost to reach 2-3\% target accuracy compared to other ML approaches. Since available materials datasets are heterogeneous providing insufficient examples for realizing the group-subgroup structure, we present the ``FriezeRMQ1D'' dataset with 8393 Q1D organometallic materials uniformly distributed across 7 frieze groups. Furthermore, by comparing the performances of FCHL and 1-hot representations, we show GS-ML to capture subgroup information efficiently when the descriptor encodes structural information. The proposed approach is generic and extendable to symmetry abstractions such as spin-, valency-, or charge order.
\end{abstract}


\maketitle
\section{\label{Introduction}Introduction}
The upsurge in the use of machine learning (ML) modeling in computational chemistry or materials science is because ML models, once trained sufficiently well on computed properties, deliver accurate new predictions at a cost lower than that of the reference method by orders of magnitude\cite{faber2017prediction,schmidt2019recent}. Developments of such novel, data-driven methods are fueled by pioneering efforts in designing and generating big data using high-throughput computation\cite{hachmann2011harvard,ramakrishnan2014quantum,chakraborty2019chemical,kirklin2015open,curtarolo2012aflowlib,jain2013commentary,kim2018polymer}. While for molecules such campaigns aim at a complete coverage of synthetically feasible chemical compound space\cite{ruddigkeit2012enumeration}, for materials, data generation is inspired by experimentally known structures\cite{hellenbrandt2004inorganic}.

In this study, we propose the group-subgroup machine learning (GS-ML) approach for
 modeling on materials datasets containing multiple crystal structures for a given stoichiometry. 
Landau applied group theory for understanding continuous phase transition of a material from a phase in high-symmetry space group, $\mathcal{G}_h$, to a phase in low-symmetry space group, $\mathcal{G}_l$, 
that is a subgroup, $\mathcal{G}_l \subset \mathcal{G}_h$\cite{stokes1984group}. Landau theory states that symmetry breaking occurs through a collective variable that transforms according to a single irreducible representation of $\mathcal{G}_h$\cite{landau1937theory}. Systematic development of structural relationships between different crystal structures was done by B\"arnighausen using crystallographic {\it group–subgroup relations}\cite{muller2013symmetry}. These relationships are
represented as a graphical tree, (B\"arnighausen-tree or $\mathcal{B}$-tree), with the most symmetric space group placed at the top.  A $\mathcal{B}$-tree not only encodes group–subgroup relations but it also correlates the Wyckoff positions during symmetry reduction\cite{muller2013symmetry}. The so-called {\it translationengleiche} subgroups ($t$-subgroups) are related to their supergroup by a decrease in the point group symmetry of the lattice while conserving the translation symmetry. A group can also have
{\it klassengleiche} subgroups ($k$-subgroups) retaining the 
point group symmetry {\it i.e.,} the crystallographic class, 
while compromising on the translation symmetry\cite{deonarine1983group,muller2013symmetry}. 
A material's thermodynamic stability that largely arises from intra-unit cell atomic interactions in a group can heuristically be understood as similar in its $t$-subgroups and $t$-supergroups. 
\begin{figure*}[!hptb]
    \centering
    \includegraphics[width=\textwidth, keepaspectratio]{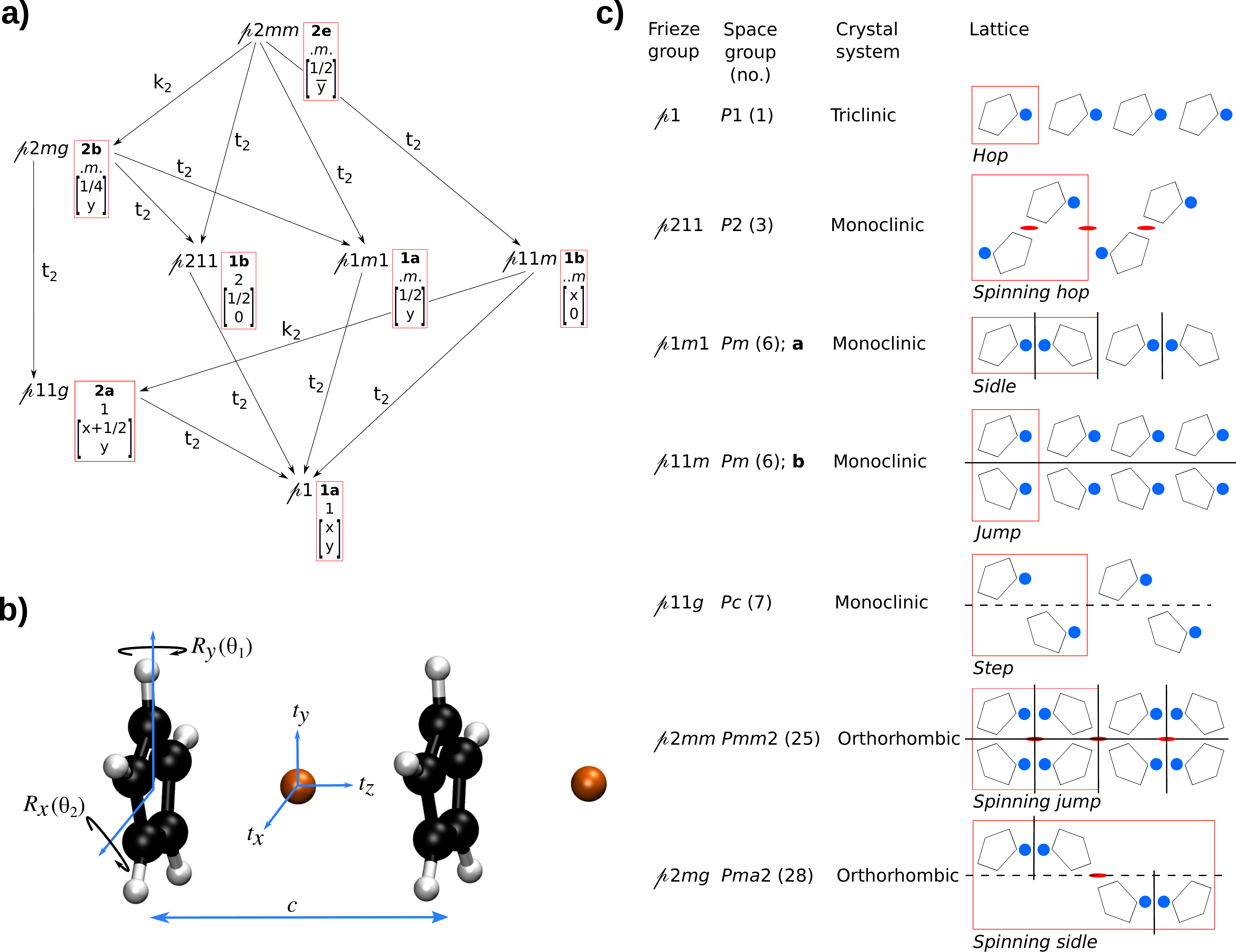}
    \caption{a) $\mathcal{B}$-tree for 7 frieze groups. The Wyckoff position occupied by the unit cell of a frieze group is shown in the box next to its name. {\it Klassengleiche} subgroups of the frieze groups are linked with an arrow marked with k$_2$ and the {\it translationengleiche} subgroups are marked t$_2$.
    b) Definitions of 6 degrees of freedom that are relaxed during the 
constrained optimization of FriezeRMQ1D materials: $\theta_1$ and $\theta_2$ are angles 
through which the ring is rotated around $x$ and $y$ axes, respectively; ${t_x,t_y,t_z}$ are
components of the translation vector of the Cu atom; $c$ is the lattice constant. 
c) Seven crystallographic frieze groups and their lattice arrangements.   
    Atoms in the solid red box comprise the unit cell. Common names of 
    frieze groups are stated below the unit cell. 
    Mirror planes are shown in solid lines while glide planes in dashed lines, 
    and a red oval symbol signifies a 2-fold rotation axis. 
    For all frieze groups, the corresponding space group names, numbers, 
    and crystal systems are also given.}
    \label{BTree}
\end{figure*}

\begin{table*}
\begin{threeparttable}
\caption{\label{tab:table3}Symmetry operations for  generating the atomic coordinates for
various formula units in a unit cell of FriezeRMQ1D materials
for seven frieze group symmetries:
$\hat{T}_j(\xi)$ denotes translation along the $j$-axis through a distance $\xi$,
$\hat{R}_x(\pi)$ denotes two-fold rotation w.r.t. the $x$-axis,
$\sigma_{ij}$ denotes mirror planes and
$\{r_k^{J}\}$ denote the set of atomic coordinates belonging to the $J-$th unit.
Each formula unit contains a five-ring and a metal.
Also given are the number of degrees-of-freedom for the crystal structure
relaxation ($N$). In all cases, the six internal degrees of freedom within a unit is
defined in Figure~\ref{BTree}b.
}
\begin{ruledtabular}
\begin{tabular}{llllll}
Group   & $N$ & \multicolumn{4}{l}{Wyckoff positions of formula units}\\
\cline{3-6}
      & &  ($x,y,z$) & ($x,\overline{y},\overline{z}$) & ($x,\overline{y},z$)${\tnote{a}}$ & ($x,y,\overline{z}$)$\tnote{b}$\\
\hline
{\calligra p}$1$    & 6 &$\{{r_k^{\rm I}}\}$ = $\{{r_k}\}$& & &\\
{\calligra p}$211$  & 6+2 &$\{{r_k^{\rm I}}\}$ = $\hat{T}_y(\xi_1)\hat{T}_z(\xi_2)$ $\{{r_k}\}$ & $\{{r_k^{\rm II}}\}$ = $\hat{R}_x(\pi)$ $\{{r_k^{\rm I}}\}$& &\\
{\calligra p}$1m1$  & 6+1 &$\{{r_k^{\rm I}}\}$ = $\hat{T}_z(\xi)$ $\{{r_k}\}$ & & &
      $\{{r_k^{\rm II}}\}$ = $\hat{\sigma}_{xy}$ $\{{r_k^{\rm I}}\}$\\
{\calligra p}$11m$  & 6+1 &$\{{r_k^{\rm I}}\}$ = $\hat{T}_y(\xi)$ $\{{r_k}\}$ & &
      $\{{r_k^{\rm II}}\}$ = $\hat{\sigma}_{xz}$ $\{{r_k^{\rm I}}\}$&\\
{\calligra p}$11g$  & 6+1 &$\{{r_k^{\rm I}}\}$ = $\hat{T}_y(\xi)$ $\{{r_k}\}$ & &
      $\{{r_k^{\rm II}}\}$ = $\hat{\sigma}_{xz}\hat{T}_z(c/2)$ $\{{r_k^{\rm I}}\}$&\\
{\calligra p}$2mm$  & 6+2 &$\{{r_k^{\rm I}}\}$ =
$\hat{T}_y(\xi_1)\hat{T}_z(\xi_2)$ $\{{r_k}\}$ &$\{{r_k^{\rm II}}\}$ = $\hat{R}_x(\pi)$ $\{{r_k^{\rm I}}\}$&$\{{r_k^{\rm \rm III}}\}$ = $\hat{\sigma}_{xz}$ $\{{r_k^{\rm I}}\}$  & $\{{r_k^{\rm \rm IV}}\}$ = $\hat{\sigma}_{xy}$ $\{{r_k^{\rm I}}\}$ \\
{\calligra p}$2mg$  & 6+2 &$\{{r_k^{\rm I}}\}$ =
$\hat{T}_y(\xi_1)\hat{T}_z(\xi_2)$ $\{{r_k}\}$ &
$\{{r_k^{\rm II}}\}$ = $\hat{R}_x(\pi)$ $\{{r_k^{\rm I}}\}$&
$\{{r_k^{\rm III}}\}$ = $\hat{\sigma}_{xz}\hat{T}_z(c/2)$ $\{{r_k^{\rm I}}\}$
&
$\{{r_k^{\rm IV}}\}$ = $\hat{\sigma}_{xy}\hat{T}_z(c/2)$ $\{{r_k^{\rm I}}\}$
\\
\end{tabular}
\end{ruledtabular}
\begin{tablenotes}\footnotesize
\item [a] for {\calligra p}$11g$ and {\calligra p}$2mg$
the Wyckoff position is ($x,\overline{y},z+1/2$).
\item [b] for {\calligra p}$2mg$
the Wyckoff position is ($x,y,\overline{z}+1/2$).
\end{tablenotes}
\label{GeometryOptTable}
\end{threeparttable}
\end{table*}

To reach a given target accuracy, training over
materials within a single space group has been shown to require a smaller training set  compared to training over materials across multiple space groups\cite{faber2016machine}. 
Yet, the application of symmetry-stratification in ML has been mostly
confined to molecular potential energy surface modeling\cite{behler2007representing}. 
This limited exploitation
of symmetry in ML modeling of molecular energetics is due to the fact that across stoichiometries, most of the information about the total electronic energy is encoded in atom-in-molecule-based fragments\cite{huang2020quantum} that seldom show any correlation with the global symmetry of a molecule. 
When modeling materials in a space group with a large unit cell, 
the proposed GS-ML approach facilitates both a reduction in the training set size and instantaneous generation of the descriptor for a {\it new} query. 
It must be noted that the number of atoms in a material's unit cell is independent of the order of the crystallographic space group\cite{mehl2017aflow,hicks2019aflow,hicks2020aflow}. 
Consider structures in the widely studied perovskite family ABX$_3$.
The most symmetric structure with the smallest unit cell is cubic in the space group, $Pm{\overline 3}m$. A simple mechanism
of tilting the rigid octahedral units results in structures
belonging to 15 subgroups with larger unit cells\cite{howard1998group}. A counterexample is the Peierls transition in a 1D chain of H-atoms. Here, the uniformly
distributed chain belongs to the $P1$ group with one atom in the unit cell.
Following Peierls distortion, the unit cell contains a H$_2$ molecule
and the system is in $P{\overline 1}$ space group. Hence, the high-symmetry
phase contains fewer atoms in the unit cell {\it only} when the phase is more compact (smaller lattice constants) than the low-symmetry phase.

The key features of GS-ML are:  1) Modeling on a class of materials belonging to a crystallographic space group with large unit cells requires the training set to include examples from smaller unit cell phases. For conventional unit cells, moving from a group to its $t$-subgroup decreases the unit cell size by a factor, $n$, which is the ratio between orders of the group and its subgroup. 
2) Querying on large unit cell materials requires a prototype descriptor made of equilibrium geometries of smaller unit cells. 
For this, one applies the splitting pattern of the Wyckoff positions (listed in the $\mathcal{B}$-tree) during symmetry lowering from $\mathcal{G}_h$ to $\mathcal{G}_l$.  Subgroup-based 
unit cell geometries also ensure the descriptor size to be 
homogeneous for establishing a faithful correlation with 
size-intensive target quantities such as 
cohesive, atomization, or formation energies per atom.

\section{\label{sec:methods}Data}

\subsection{\label{sec:frieze}Ring-metal Q1D materials in seven crystallographic frieze groups: FriezeRMQ1D}
We design a dataset that consists of Q1D materials with a ring-metal pair as the formula unit. In a previous study we reported geometries, electronic and phonon properties 
of these materials composed of 
11 monovalent metals: Na, K, Rb, Cs, Cu, Ag, Au, Al, Ga, In, and Tl combined with
109 heterocyclic rings\cite{kayastha2021high}. The ring-metal Q1D dataset (RMQ1D) contains
1199 materials, where the rings are 
generated by combinatorial substitution of C atoms in the cyclopentadienyl (Cp) anion with
B, N, or S atoms of all possible valencies. The geometries of RMQ1D materials were
generated without enforcing any symmetry restrictions, hence these materials are with the
$P$1 space group symmetry. 
While, in principle, it is possible to design crystallographic
prototypes in all 230 space groups, here we restrict our exploration to a smaller subset
corresponding to 7 frieze groups.
In a frieze pattern, commonly encountered in architecture or textile design, a two-dimensional unit is repetitive in one direction. Limiting the symmetry coverage to frieze groups preserves the low-dimensional nature of the RMQ1D materials.
A frieze group is a set of all elements that define the symmetries of a frieze pattern, namely, an axis of translation, a two-fold rotation axis, 
mirror planes, and a glide plane. The $\mathcal{B}$-tree of frieze groups showing group-subgroup relationships is presented in Figure~\ref{BTree}a. For every frieze group, there is an isomorphic crystallographic space group.

RMQ1D materials belonging to the
smallest frieze group symmetry, {\calligra p}$1$, contain repeated
ring-metal units. 
In an earlier study, we reported properties of 1199 RMQ1D materials based on full geometry relaxations\cite{kayastha2021high}. Exploring the complete materials space in 7 frieze groups
warrants careful consideration of internal degrees of freedom in the ring-metal unit for
geometry relaxations. 
For this purpose, we have defined the most relevant internal
coordinates in Figure~\ref{BTree}b. As shown in Figure~\ref{BTree}c, in the 6 larger frieze groups, the FriezeRMQ1D materials contain more than one ring-metal formula unit. Hence, for their minimum energy crystal structures 
to preserve the frieze group symmetries and not
result in low-symmetry structures with larger unit cells, 
symmetry constraints are inevitable\cite{lenz2019parametrically}.
Such constrained optimizations can be performed with the knowledge of symmetry operations relating the atomic coordinates in crystallographic Wyckoff positions. For the seven frieze groups, we
summarize these relationships in Table~\ref{GeometryOptTable}. 
Groups {\calligra p}$211$, {\calligra p}$1m1$,  {\calligra p}$11m$, and {\calligra p}$11g$
contain two Wyckoff positions, while
the larger, high-symmetry groups {\calligra p}$2mm$ and {\calligra p}$2mg$ contain
four Wyckoff positions. Groups {\calligra p}$1m1$,  {\calligra p}$11m$, and {\calligra p}$11g$ provide
the flexibility to relax the unit cell dimension through the translation variable, $\xi$. The remaining groups 
containing a 2-fold rotation, {\calligra p}$211$, {\calligra p}$2mm$ and {\calligra p}$2mg$
require two translation variables, $\xi_1$ and $\xi_2$, to be optimized in addition to the 6 intra formula 
unit coordinates defined in Figure~\ref{BTree}b. 
Schematic structures of FriezeRMQ1D materials in seven frieze
phases resulting from an application of the transformations presented in 
Figure~\ref{BTree}b and Table~\ref{GeometryOptTable}, are summarized in 
Figure~\ref{BTree}c.

\begin{figure*}[!hpt]
    \centering
    \includegraphics[width=\textwidth,  height=\textheight, keepaspectratio]{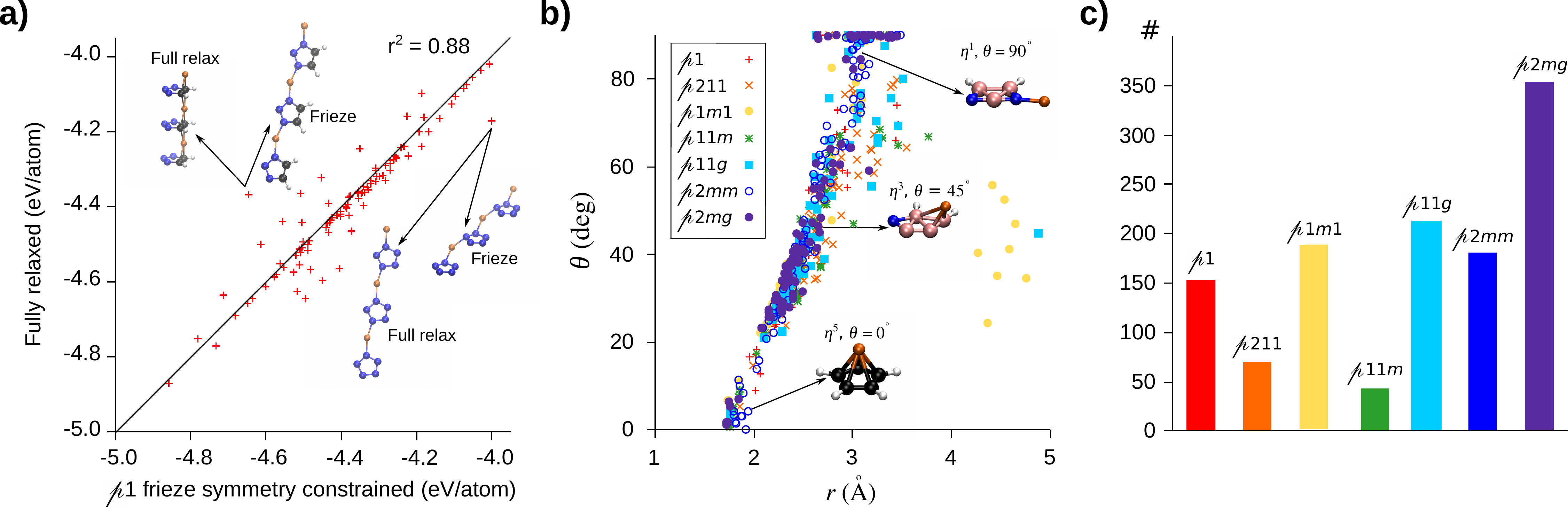}
    \caption{Structure-property diversity in the FriezeRMQ1D dataset:
    a) Effect of geometry optimizations (full vs. frieze symmetry constrained relaxations) on atomization energies of materials 
    containing a ring-Cu unit. 
    Structures of entries
    with large deviations from the $y=x$ line are shown.
    b) Preference for hapticity ($\eta$) of intra formula unit 
    bonding is shown through joint variation of the ring-metal
    distance, $r$ (in \AA), and the ring-slip angle, $\theta$ (in degrees).
    c) Distribution of the thermodynamically most stable phase
    across 7 frieze groups. For each stoichiometry, the most stable phase is
    considered.
    }
    \label{fig:rtheta}
\end{figure*}

Several analyses can be performed to understand the diversity of the FriezeRMQ1D dataset containing
$7\times1199=8393$ Q1D materials. Here, we briefly comment on the important
aspects relevant to GS-ML modeling in this dataset. Key geometric and energetic
trends in the FriezeRMQ1D dataset are illustrated in Figure~\ref{fig:rtheta}. 
Firstly, it is important to compare the thermodynamic stabilities of the materials in
constrained relaxations with frozen rings to those from more accurate, full relaxations. 
Figure~\ref{fig:rtheta}a presents the atomization energies of the {\calligra p}$1$ subset
with Cu metal. As a general trend, we note that atomization energies of constrained geometries approach the fully relaxed values. For a majority of the systems, a preference for fully relaxed structure has shown small energetic preference indicated by the points accumulated below the $y=x$ line. Nearly a couple dozen systems show large energy deviations due to conformational or $\sigma$-vs.-$\pi$ bonding preferences. Overall, a value of 0.88 for the Pearson correlation coefficient, $r^2$, implies the geometries of the FriezeRMQ1D materials presented here to be good approximations to the fully relaxed ones reported in \RRef{kayastha2021high}. Depending on the ring-metal combinations and frieze symmetries, the formula units prefer
bonding patterns of different hapticities as shown in Figure~\ref{fig:rtheta}b. Structures showing
$\pi$-bonding with $\eta^5$ bonding pattern exhibit short distances, $r$, between the metal and the center of the rings. With an increase in $\sigma$-character, $r$ increases along with a rise in the
ring-slip angle, $\theta$.
Fully $\sigma$-bonded structures with $\eta^1$ feature on the 
top of the image corresponding to $r\ge3{\rm \AA}$ and $\theta \approx 90^\circ$. 
A preference for a co-planar arrangement of the ring-metal pair
coincides with an increase in $r$. 
A few {\calligra p}$1m1$ structures 
show $r>4{\rm \AA}$ indicating large repulsive interactions between the formula units. Such systems show an overall
thermodynamic preference for other frieze group symmetries. For each composition in the dataset, the energetically preferred phase is summarized in Figure~\ref{fig:rtheta}c. Out of 1199 stoichiometries, about 150 crystallize in 
{\calligra p}$1$ phase, while the least and most populated phases are {\calligra p}$11m$ and {\calligra p}$2mg$, respectively.


\subsection{\label{sec:dft}Electronic structure calculations}

Symmetry constrained geometry relaxations 
were performed with 
the all-electron, numerical atom-centered orbital code
FHI-aims\cite{blum2009ab} with the  Perdew-Burke-Ernzerhof (PBE)\cite{perdew1996generalized} functional.
Since the goal of the present study is to explore the capabilities of the ML approach, 
rather than presenting a robust database for benchmarking the first-principles method with experiment, 
geometry optimizations
were performed only for the 763 Cu-systems (109 rings combined with 7 frieze groups). Energies of the materials
with the remaining 10 metals were calculated in a single point fashion using geometries of the 
Cu-based materials.
In all calculations, we used
$1\times1\times 64$ $k$-grids, and a tight/tier-1 basis set.
Search for a stationary point on the potential energy surface was 
performed using the 
Broyden-Fletcher-Goldfarb-Shannon 
(BFGS)\cite{broyden1970convergence,fletcher1970new,goldfarb1970family,shanno1970conditioning}
minimization procedure using the PBE\cite{perdew1996generalized} total energy.
In the BFGS procedure we used a convergence thresholds of $10^{-4}$ $e/{\rm \AA}$ for the
gradient norm and $10^{-6}$ $e/{\rm \AA}^3$ for the electron density.
The impact of the frieze group arrangement on the internal structure
of the ring-metal formula unit is  illustrated in Figures~\ref{fig:rtheta}b and \ref{fig:rtheta}c.  
Symmetry constrained geometry relaxations have been performed by mapping the 
relevant atomic and lattice degrees of freedom to a small set of parameters
that can be unconditionally relaxed\cite{lenz2019parametrically}. For the FriezeRMQ1D materials,
the definitions of these parameters are collected in Table~\ref{GeometryOptTable}.
It must be noted that all the lattice vectors are freely relaxed, hence 
the volume of the unit cell does not remain fixed in the symmetry constrained calculations.
Single point PBE0\cite{adamo1999toward} calculations were performed at relaxed 
geometries for accurate estimation of energies and band gaps.
The total computational cost for PBE-level symmetry constrained optimizations and PBE0-level single-point energy evaluations are 14 CPU months, and 45 CPU days, respectively.

\section{Methodology}
\subsection{Representations for the FriezeRMQ1D dataset}
In this study, we explore the suitability of 
a structure-based descriptor and a composition-based one to use for GS-ML. 
We compare and contrast their
performances in terms of ML-cost and accuracy. 
The Faber--Christensen--Huang--Lilienfeld (FCHL)\cite{faber2018alchemical,christensen2020fchl} descriptor is one of the best structure-based descriptors. FCHL includes up to
three-body interaction terms and also accounts for alchemical
variations for modeling across stoichiometries. One of the prerequisites of using the FCHL representation is the availability of appropriate structures. In the context
of GS-ML methodology, the goal is to model properties of materials in a 
large unit cell phase. 
However, when minimum energy geometries are required for both training and 
out-of-sample predictions, the total ML cost equals that of the cost of DFT
calculation for the entire dataset, offering no practical advantage of ML modeling.
Hence, for modeling materials in a particular frieze group, we use 
the equilibrium geometries of the corresponding materials in a 
smaller unit cell phase, {\it i.e.,} the ML modeling is done using the subgroup geometry and building a supergroup descriptor by applying the respective group operations. 



The need for accurate subgroup geometries can further be reduced by using 
a fingerprint-\cite{meldgaard2020structure,huan2015accelerated,batra2019general,von2015fourier,imbalzano2018automatic}, 
atom-level-\cite{de2016comparing}, or
composition-based representations \cite{ward2017including} that do not 
require structural information for out-of-sample predictions. In a previous study\cite{faber2016machine}, a fingerprint descriptor has delivered more accurate ML predictions than the FCHL formalism for modeling the energetics of 2 million elpasolites. For each elpasolite, the fingerprint descriptor uniquely maps to the material's
composition. On the other hand, for ML modeling on a structurally diverse materials
dataset such as the
open quantum materials database (OQMD)\cite{kirklin2015open}, a 
descriptor containing structural information offered good prediction accuracy. 
A subset of OQMD, that is reported in the
inorganic crystal structure database (ICSD)\cite{hellenbrandt2004inorganic} contain only a few entries of multiple structures for the same chemical formula (hence compositionally
dominant). For ML modeling
on this ICSD subset of OQMD, 
augmenting structural information to a composition-based fingerprint 
one did not improve the prediction accuracy\cite{ward2017including}.

\begin{figure}[!hpt]
    \centering
    \includegraphics[width=8.5cm]{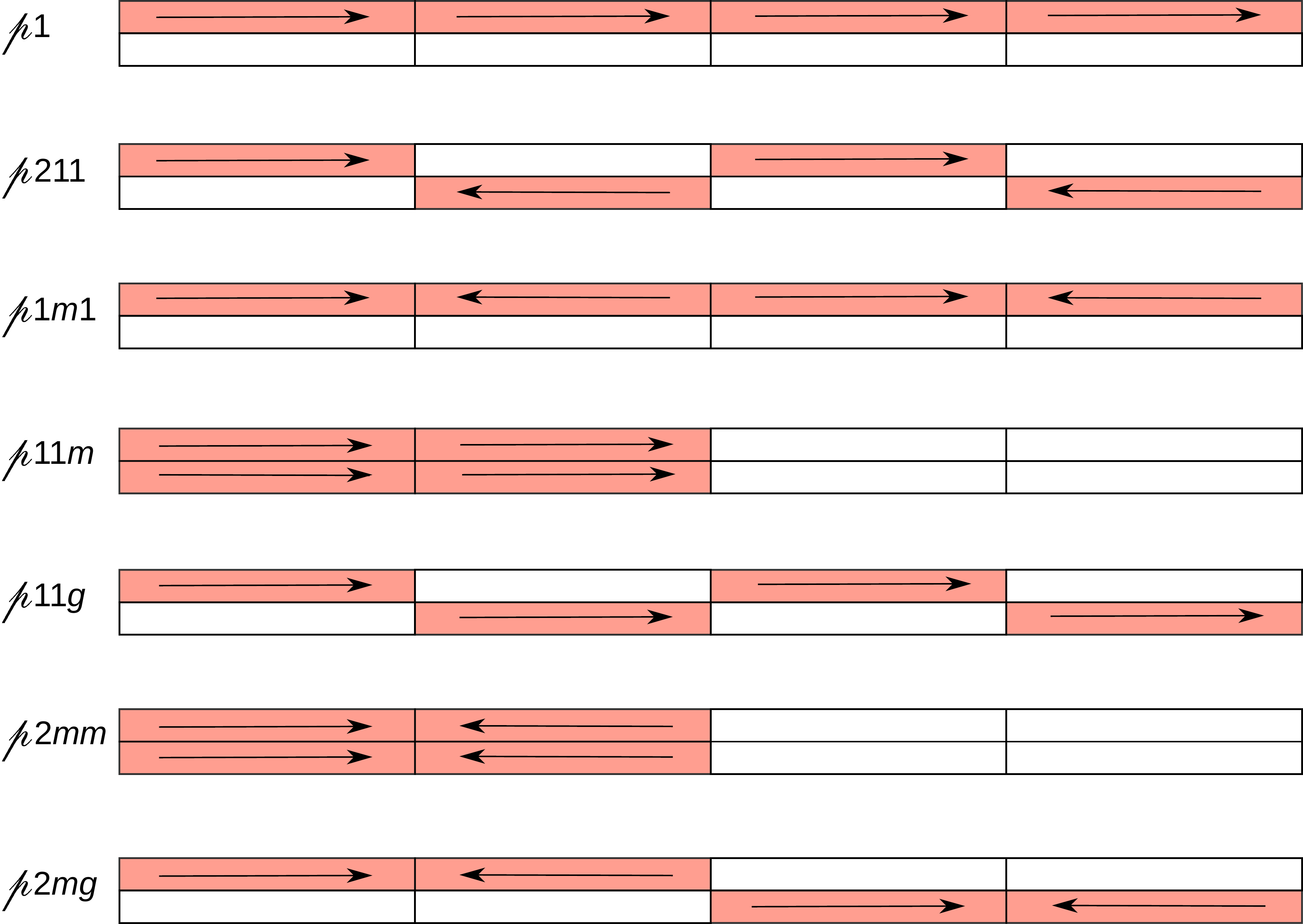}
    \caption{Encoding of Wyckoff positions in the 1-hot representations for 
    FriezeRMQ1D materials. Each shaded rectangle is a 19-bit integer
    encoding the composition of the ring-metal unit. 
    Empty rectangles contain
    zeroes for padding. The direction of the arrows signifies the bit sequence.
    }
    \label{wyckoffencoding}
\end{figure}

The FriezeRMQ1D dataset presented in this work is both structurally and compositionally rich. Hence, it is of interest to test the performance of a descriptor devoid of the materials' three-dimensional structural information.  
To this end, we explore a binary fingerprint representation---the 1-hot vector---that encodes the stoichiometry of the materials and Wyckoff positions of the 7 frieze groups.
The 1-hot vector representation of the FriezeRMQ1D materials encodes information about the unit cell composition and crystallographic Wyckoff positions (Fig.~\ref{wyckoffencoding}). 
This representation is made commensurate across frieze groups by considering a minimum of 8 formula units to apply all possible group operations. The size of the vector per formula unit is 19 bits, of which, the first 15 are reserved for the sites in the 5-ring---each site requires 3 bits to store the 8 possible atom-valency combinations: C (000), CH (001), B (010), BH (011), N (100), NH (101), S (110), and SH (111)---while the last 4 bits are required to store 11 metal types Na (0000), K (0001), Rb (0010), Cs(0011), Cu (0100), Ag(0101), Au (0110), Al (0111), Ga (1000), In (1001), Tl (1010). For example, the 1-hot vector for Cp-Na is 0010010010010010000, which represents 
a shaded rectangle with a
right-pointing arrow in Figure \ref{wyckoffencoding}.

The scope of the problem presented here is not limited 
by the representation, hence, one can also apply other descriptors such 
as 
row-sorted Coulomb matrix \cite{rupp2012fast},
bag-of-bond (BoB), \cite{hansen2015machine},
SLATM \cite{huang2020quantumm},
MBTR\cite{langer2020representations}, or SOAP\cite{de2016comparing} to exploit group-subgroup structural relationships in ML modeling of materials.


\subsection{Kernel-ridge regression}
We use kernel-ridge regression\cite{scholkopf2002learning,rupp2012fast} for modeling atomization energies, $E$, of the FriezeRMQ1D materials. The energy of a query material, $q$, is estimated as the linear combination of kernel functions, each centered on a 
training material, $t$.  The
kernel functions take as argument the similarity between the query and the 
training materials quantified through a descriptor, ${\bf d}$
\begin{equation}
E^{\rm est}({\bf d}_q) = \sum_{t=1}^N c_t k({\bf d}_q, {\bf d}_t).
\label{eq:krr1}
\end{equation}
Here, we use a Laplacian kernel, 
$k({\bf d}_q, {\bf d}_t)=\exp(-|{\bf d}_q - {\bf d}_t|/\sigma)$, where
 $\sigma$ defines the length scale of the kernel function. The
 fit-coefficients, $c_t$, are obtained by solving the linear system
\begin{equation}
\left[ {\bf K}+\lambda {\bf I} \right]\ {\bf c} = {\bf E}^{\rm DFT}
\label{eq:krr3}
\end{equation}
In all ML calculations, we use a fixed-value for the 
regularization strength, $\lambda=10^{-4}$, as a preconditioning measure. 
For determining a suitable kernel-width, we followed the ``single-kernel recipe''
proposed in \RRef{ramakrishnan2015many}
\begin{equation}
\sigma_{\rm opt}^{\rm max} = d_{ij}^{\rm max}/\log(2).
\label{eq:singlewidth}
\end{equation}
where $d_{ij}^{\rm max}$ is the 
largest descriptor difference among training entries. Use of fixed hyperparameters
${\lambda,\sigma}$ enable rapid training, facilitating adequate 
shuffling of the training set to prevent any bias. 

For this dataset with the fingerprint descriptor, fixed values of 
$\sigma=92.3$ and $\lambda=10^{-4}$, result 
in  an out-of-sample error of 0.099 eV/atom for a training set size of 500.
We have tested the 
performance of this single-kernel {\it Ansatz} by comparing with 
a cross-validated model for 1199 materials in the {\calligra p}1
frieze group. 
When these hyperparameters are optimized through a 5-fold cross-validation,
the out-of-sample error over 699 entries drops to 0.093 eV/atom 
for  $\sigma=574.0$ and $\lambda=2\times 10^{-4}$. The resulting gain in accuracy is negligible compared to the residual uncertainties due to shuffles. 
For the FCHL formalism, we have 
determined an optimal kernel width of $\sigma=5$ through scanning with a fixed regularization strength of $\lambda=10^{-4}$. 
A large cutoff distance of 15 \AA was used to capture
the global structure of the unit cell. 
This descriptor has a tensor structure and a 
direct evaluation of the
kernel matrix elements is the preferred approach
as implemented in the program QML\cite{QML}.

\section{Results and Discussions}
\begin{figure*}
    \centering
    \includegraphics[width=0.95\textwidth]{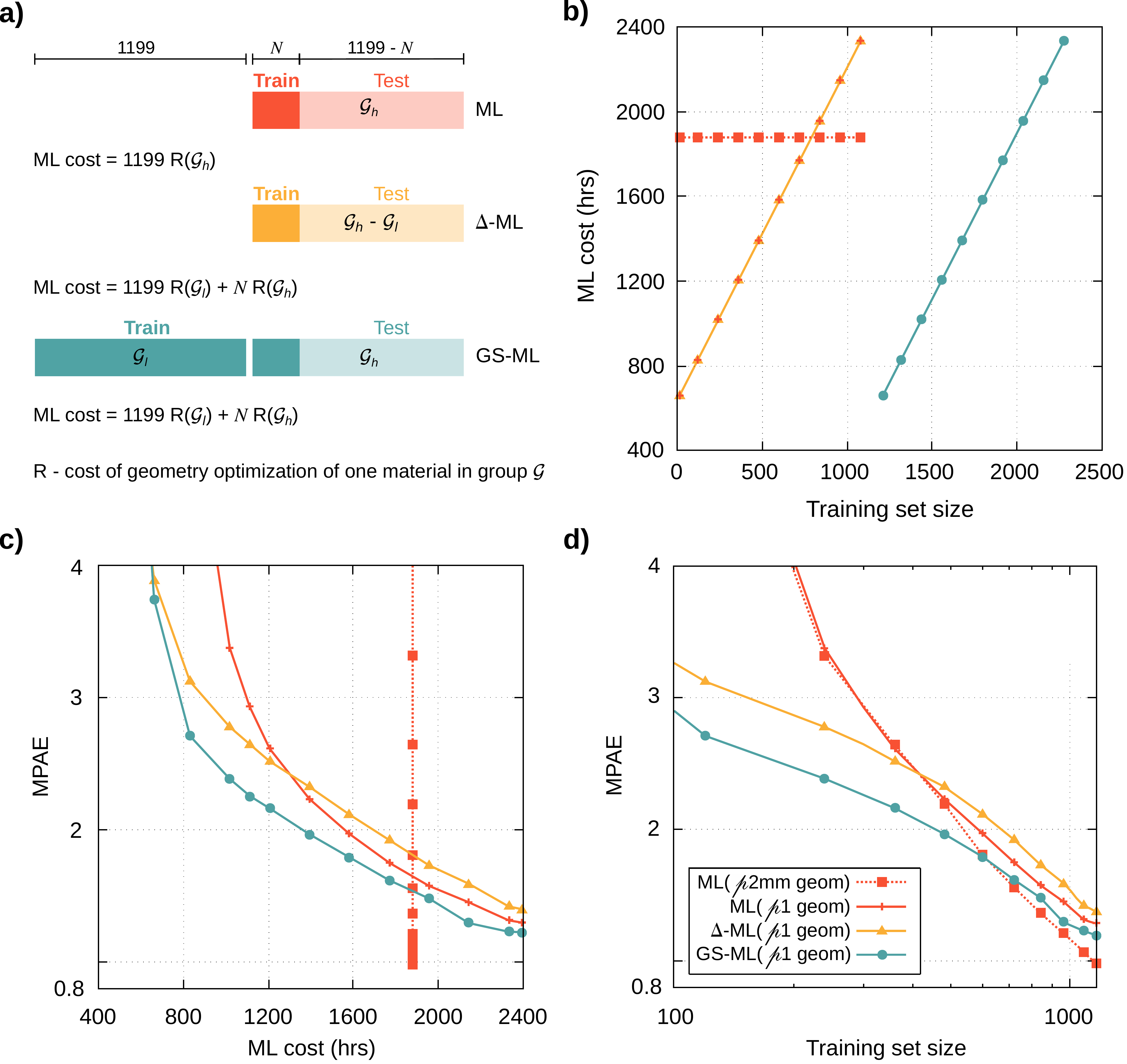}
    \caption{Comparison of ML, $\Delta$-ML and GS-ML methods
    with the FCHL formalism.
    a) Definition of test-train splits for out-of-sample 
    predictions of materials in the supergroup, $\mathcal{G}_h$. 
    The ML model is trained on atomization energies $E(\mathcal{G}_h)$ while
    the $\Delta$-ML model is trained on difference in atomization energies
    $\Delta=E(\mathcal{G}_h)-E(\mathcal{G}_l)$. 
    In GS-ML model is trained on $E(\mathcal{G}_h)$ with
    all examples from the subgroup $\mathcal{G}_l$
    included in the training set. 
    ML cost for each ML model is given as a function of geometry optimization
    in $\mathcal{G}_l$ and $\mathcal{G}_h$.
    b) ML cost (in CPU hours) vs. training set size.
    ML modeling is done with two choices of geometries (red solid/plus for prototype {\calligra p}$2mm$ geometries made of fragments from relaxed {\calligra p}1 geometries and
    red dotted/square for relaxed {\calligra p}$2mm$ geometries).  
    Both $\Delta$-ML (yellow/triangle) and GS-ML (green/circle) models use 
    prototype {\calligra p}$2mm$ geometries. Cost of all three ML models 
    using {\calligra p}1 geometries are the same.
    For colour coding see panel-d. 
    c) Out-of-sample mean percentage 
    absolute error (MPAE) is plotted as a function of ML cost. 
    Results are based on mean values over 20 shuffles. 
    For colour coding see panel-d. 
   d) Learning curves for ML, $\Delta$-ML and GS-ML methods. For GS-ML, the
    training set size is shifted by 1199 for a fair comparison of the
   model's cost and accuracy with that of $\Delta$-ML's. 
    }
    \label{ml}
\end{figure*}

For ML modeling of materials with large unit cells, the GS-ML approach offers the best possible
cost-to-accuracy trade-off. In this study, we present a proof-of-the-concept by applying GS-ML for modeling the
atomization energies of the FriezeRMQ1D materials and compare 
the results to conventional ML modeling.  
Traditional ML modeling involves training on the relaxed geometries with target properties evaluated in these geometries. To remove any sampling bias, the dataset is 
shuffled and distributed over test/train sets. 
Figure \ref{ml}a shows the cost of ML, size of the dataset, and train-test split for predicting materials in $\mathcal{G}_h$.
For the FriezeRMQ1D materials, $\mathcal{G}_h$ structures are with large unit cells while the structures in $\mathcal{G}_l$
are with small unit cells. It is common in ML modeling of molecular or materials properties
to use structure-based
descriptor derived at relaxed equilibrium geometries. This requires geometry optimizations of all
entries in the training set and geometries at the same level are also 
required for querying. 
Hence, as a function of training set size, the ``ML cost'' remains constant as shown in 
Figure \ref{ml}b. In the following, we consider the cost of the linear algebra procedures involved
in training the ML model negligible compared to the cost of data generation. 

Symmetry correlation of materials across phases, as encoded in the
crystallographic $\mathcal{B}$-tree, 
can be used to decrease the cost of descriptor generation for $\mathcal{G}_h$ (i.e. large unit cell) phases. 
This can be achieved by using the subgroup geometry in $\mathcal{G}_l$ (with fewer formula units) and the splitting of Wyckoff sites. To achieve this
one must choose appropriate values for the internal coordinates $\xi$ defined in Table~\ref{GeometryOptTable}. For modeling atomization energies of materials in the {\calligra p}$2mm$ phase, we use the equilibrium structures in the subgroup {\calligra p}1. For example, the unit cell of Cp-Na in the {\calligra p}1 frieze group contains 11 atoms while that in the
{\calligra p}$2mm$ frieze group contains 44 atoms, see Figure \ref{BTree}c for more details on the number of formula units in the unit cell of a frieze group. Using the {\it subgroup geometry}, one can perform 
ML modeling in three different ways: 
\begin{enumerate}
    \item ML (subgroup geometry) -- with descriptors based on the subgroup geometries.
    \item $\Delta$-ML (subgroup geometry) -- with descriptors based on the subgroup geometries and using the atomization energies of the subgroup phase as a baseline.
    \item GS-ML (subgroup geometry) -- where all the examples from the subgroup are kept in the training set and the descriptors for materials in $\mathcal{G}_h$ are based on the subgroup geometries. 
\end{enumerate}
All the stated ML recipes depend on subgroup geometries and geometry optimization in the expensive $\mathcal{G}_h$ phase
is required only for calculating 
the atomization energies of the training examples. Hence, these three ML procedures exhibit the same cost for data generation (see Figure \ref{ml}b). Note that GS-ML requires a larger, symmetry-stratified, training set than the other approaches while exhibiting the same {\it total} ML cost. 


The main reason for the success of structure-based ML approaches such as the FCHL formalism is that the design of the descriptors closely follows the assumptions involved in the potential energy surface modeling. The FCHL formalism
utilizes radial basis functions for radial and angular coordinates, that are damped with carefully chosen weight functions. 
While no further assumptions have been made to result in biased predictions, the formalism delivers
good accuracies when trained on minimum energy geometries and for predicting
the energies calculated at these geometries. 
Figures~\ref{ml}c and \ref{ml}d present the learning curves for modeling  {\calligra p}$2mm$ energies using
 {\calligra p}$2mm$ (red/dotted, square) and subgroup ({\it i.e.,}  {\calligra p}$1$, red/solid, plus) geometries. At the limit of large training set sizes (hence greater cost) one notes the equilibrium geometry-based ML to result in a lower mean percentage absolute error (MPAE). A similar trend has also been noted in a recent work using ML for predicting molecular and materials 
 geometries that in turn can be used for
instantaneous generation of descriptors\cite{lemm2021energy}. 

 $\Delta$-ML has been previously shown to improve the performance of ML\cite{ramakrishnan2015big, ramakrishnan2015electronic, gupta2020revving}. 
 In this approach an inexpensive baseline quantity (often the same target quantity calculated with a cheaper model)
 can be subtracted from the target, and the ML model is trained only on the $\Delta={\rm target}-{\rm baseline}$. For out-of-sample predictions, the ML model estimates the $\Delta$ to which the baseline value is added to approximate the target quantity. In this study, we use atomization energies of the subgroup structures as the baseline. 
 For FriezeRMQ1D materials, the difference between {\calligra p}$2mm$ and {\calligra p}$1$ atomization energies amounts to packing interaction between the formula units while 
intra-formula unit contributions cancel. 
For reaching a target accuracy of 2.5-3\%, $\Delta$-ML
offers a better cost-to-accuracy trade-off compared to direct-ML 
as shown in Figures~\ref{ml}c and \ref{ml}d. 
For lower accuracies, the direct ML
method delivers more favorable learning rates. We have also tried another baseline, DFT energies of the {\calligra p}$2mm$ phases
calculated afresh with the subgroup geometries and noted a similar learning curve. 
When ML-training is saturated with the chemical physics encoded in the baseline, both $\Delta$-ML and (direct) ML models approach similar prediction errors\cite{ramakrishnan2015big}.
\begin{table}
    \caption{Mean percentage absolute error (MPAE) for out-of-sample predictions on materials in $\mathcal{G}_h$.
    FCHL formalism permits only {\it translationengleiche} 
    GS-ML while 1-hot representation enables training across 
    all 7 frieze groups. Numbers in bold correspond to results from ML and off-diagonal
    entries in the table correspond to GS-ML. For ML, training is done with
    25\% examples from $\mathcal{G}_h$. For GS-ML, in addition, all 
    examples from the $\mathcal{G}_l$ are kept in training.
    }
    \begin{tabularx}{\columnwidth}{l r l r r r r r r r}
    \hline
    \hline 
   \multicolumn{2}{l}{$\mathcal{G}_l$ }&~~~~~&\multicolumn{7}{l}{MPAE for predictions in $\mathcal{G}_h$} \\
    \cline{4-10}
  &    & &
  \multicolumn{1}{l}{{\calligra p}$1$} & 
  \multicolumn{1}{l}{{\calligra p}$211$} &  \multicolumn{1}{l}{{\calligra p}$1m1$} & \multicolumn{1}{l}{{\calligra p}$11m$} & \multicolumn{1}{l}{{\calligra p}$11g$} & \multicolumn{1}{l}{{\calligra p}$2mm$} & \multicolumn{1}{l}{{\calligra p}$2mg$}\\
   \hline 
     \multicolumn{10}{l}{FCHL} \\
  {\calligra p}$1$   &  & &{\bf 1.6} & 1.7  & 1.8  & 1.3  & 1.6  & 2.2  & 2.5 \\
  {\calligra p}$211$ &  & & &{\bf 2.2} & & & & 2.4 & 2.8\\
  {\calligra p}$1m1$ &  & & & &{\bf 1.9} & & & 2.0 & 2.6\\
  {\calligra p}$11m$ &  & & & & & {\bf 2.0}& & 2.5 & \\
  {\calligra p}$11g$ &  & & & & & &{\bf 2.3} &     & 3.2\\
  {\calligra p}$2mm$ &  & & & & & & & {\bf 3.0} &   \\
  {\calligra p}$2mg$ &  & & & & & & &    & {\bf 3.7} \\
  \hline
   \multicolumn{10}{l}{1-hot} \\
   {\calligra p}$1$  & &&    {\bf 3.0}     & 2.5     & 2.6    &  2.8    &  2.8    &  2.8     & 3.9    \\
  {\calligra p}$211$ & &&    2.9     & {\bf 2.5}     & 2.4    &  2.9    &  2.9    &  3.1     & 3.8    \\
  {\calligra p}$1m1$ & &&    3.0     & 2.4     & {\bf 2.5}    &  3.0    &  2.8    &  3.1     & 3.9    \\
  {\calligra p}$11m$ & &&    2.9     & 2.5     & 2.4    &  {\bf 3.0}    &  2.8    &  3.1     & 3.8    \\
  {\calligra p}$11g$ & &&    2.8     & 2.5     & 2.4    &  2.8    &  {\bf 2.9}    &  3.1     & 3.9    \\
  {\calligra p}$2mm$ & &&    3.0     & 2.5     & 2.4    &  3.0    &  2.9    &  {\bf 3.1}     & 3.9    \\
  {\calligra p}$2mg$ & &&    3.0     & 2.4     & 2.5    &  3.0    &  2.9    &  3.0     & {\bf 3.9}    \\
   \hline 
  \hline 
    \end{tabularx}
    \label{tab:my_label}
\end{table}

Compared to ML and  $\Delta$-ML, the GS-ML strategy results in the best cost-to-accuracy trade-off,
see Figures~\ref{ml}c and \ref{ml}d (green/solid, circle). 
In particular, GS-ML offers the fastest way
to reach a target accuracy of 2-3\%. 
For 1199 {\calligra p}$2mm$ materials, GS-ML requires fewer than 100 examples to predict the atomization energies of 1099 
out-of-sample examples to 3\% accuracy, corresponding to a mean absolute error of
0.12 eV/atom. 
When using subgroup geometries, GS-ML offers the best performing learning curve.
However, it must
be noted that for large training set sizes, a direct ML model based on the equilibrium {\calligra p}$2mm$ geometries delivers better results with
asymptotic prediction errors of less than 1\%. 

We searched for better group-subgroup combinations for the FriezeRMQ1D dataset and found {\calligra p}$1m1$ to deliver
a lower MPAE compared to {\calligra p}$1$ for 
out-of-sample predictions in {\calligra p}$2mm$ (see Table~\ref{tab:my_label}). 
However, it must be noted that the cost of geometry optimizations of the
{\calligra p}$1m1$ phase is greater compared to that in {\calligra p}$1$. 
The ratio of MPAE (ML) and MPAE (GS-ML) 
is maximum for the {\calligra p}$1$-{\calligra p}$11m$ combination. Direct modeling on {\calligra p}$2mg$ with 25\% 
data results in an MPAE of 3.7 which drops to 2.5\% for
{\calligra p}$1$-{\calligra p}$2mg$ GS-ML.

In contrast to FCHL, the 1-hot representation permits not only {\it klassengleiche} GS-ML but also facilitates training on large unit cell examples to predict the small unit cell materials. 
Our results show the errors of GS-ML 
with the 1-hot representation
to be similar to that of traditional ML modeling (see Table~\ref{tab:my_label}). This is because of the inherent
assumption in GS-ML which requires substructure similarity in phases between a group and its subgroup, that cannot be captured by composition-based 1-hot encoding. 

Generalizing the results discussed above,
the FCHL approach in combination with {\calligra p}$1$ data, facilitates
modeling of atomization energies in {\calligra p}$2mm$
with a prediction error of $< 3$\% using 10\% reference data (see Figure~\ref{ml}d). Although the FriezeRMQ1D materials are  low-dimensional, the CPU costs for their DFT energy evaluation is a
function of number of electrons in the unit cell. For the high-symmetry phases
with the Cp ring and Cu metal, the unit cell contains 256 electrons
making their computational complexity comparable to that of 
bulk ternary phases of perovskites or transition metal-based pnictides/chalcogenides.
The phases of these materials can also 
be related in a $\mathcal{B}$-tree providing scope for GS-ML in compounds such as perovskites\cite{steele2020phase}.
It must be noted that the cost for generating the data for several small unit cell materials is often
negligible compared to that of large unit cell ones. 
For instance, the ratio between the time taken for
single point energy calculations in 
{\calligra p}$1$ and {\calligra p}$2mm$ phases for the 
Cp-Cu system is $<0.02$. 
Modeling the energies of a few thousand ternary materials demands a data-generation cost for only a few hundred examples. 
Compared to this, the cost for performing DFT-level geometry optimizations of the entire set in a compact phase is often negligible.


\section{Conclusions}
We have presented the group-subgroup machine learning (GS-ML) formalism for efficient modeling of materials properties in complex phases with multiple formula units. The approach exploits transformation relationships between crystallographic space groups and is applicable for materials datasets where more than one crystal structures are available per stoichiometry. To provide a proof-of-the-concept for GS-ML, 
we generated a new Q1D materials dataset ``FriezeRMQ1D'' with chemical compositions spanning 11 metals, 109 rings, and 
uniformly distributed in all seven frieze groups.
For the resulting 8393 materials, minimum energy geometries of desired symmetries were 
calculated using constrained relaxations. 
The target property of interest is the atomization energy
calculated at the hybrid-DFT level. 
To facilitate further symmetry-based explorations, all data generated for this study are made publicly available. 

We designed a 1-hot vector representation containing information about the cyclic structure of the ring, its stoichiometry, metal, and the frieze group symmetry of the lattice. Besides, we have tested the performance of the FCHL formalism that captures structural and alchemical similarities.  We depend on percentage error as a reliable error metric than absolute errors because other studies have shown the prediction errors of ML models for materials to depend strongly on the dataset\cite{faber2018alchemical}. 
While the 1-hot representation did not result in improved performances in GS-ML, the representation may be utilized for 
rapid data-mining on the entire FriezeRMQ1D dataset.

We analyzed the performance of GS-ML for FriezeRMQ1D materials 
through all possible combinations of frieze groups. The proposed formalism is complementary to the $\Delta$-ML approach. 
The most attractive feature of GS-ML is that it 
alleviates an explicit dependence of materials descriptors on DFT-level
minimum energy geometries. For traditional ML modeling on the atomization energies of the 
FriezeRMQ1D materials in the {\calligra p}$2mm$ phase,
the FCHL formalism delivered predictions with $<1\%$ error when the ML model is trained
using minimum energy geometries.
For prediction errors in the range 2--3\%, the GS-ML approach 
incurs only half the cost
involved in training set generation compared to direct ML modeling.
Hence, constructing structural descriptors for materials phases with a 
complex unit cell arrangement using the geometries of 
compact phases along with the knowledge of Wyckoff positions splitting 
widens the application domain of ML for materials modeling.
 
The general conclusions drawn from our results are independent of the ML formalism and datasets. It will be interesting to see if this strategy can be
adapted to modeling economically important materials like perovskites\cite{howard1998group} or Fe-based pnictides\cite{shatruk2019thcr2si2}, the latter has received a lot of attention as superconductor candidates. Similarly,
the chemical physics common to non-magnetic and spin-collinear phases of materials can also
be exploited using group-subgroup relations in the distribution of local magnetic moments.

\section{Data Availability}
Atomic coodinates of all 8393 FriezeRMQ1D materials, 
Wyckoff-encoded 1-hot vectors, and PBE0
atomization energies are available at \href{http://moldis.tifrh.res.in/data/FriezeRMQ1D}{http://moldis.tifrh.res.in/data/FriezeRMQ1D}.
Input and output files of corresponding calculations are deposited in the NOMAD repository \href{https://dx.doi.org/10.17172/NOMAD/2021.02.13-1}{(https://dx.doi.org/10.17172/NOMAD/2021.02.13-1)}.

\section{Acknowledgments}
We acknowledge support of the Department of Atomic Energy, Government
of India, under Project Identification No.~RTI~4007. 
All calculations have been performed using the Helios computer cluster, which is an integral part of the MolDis Big Data facility, TIFR Hyderabad \href{http://moldis.tifrh.res.in}{(http://moldis.tifrh.res.in)}.

\appendix*
\section{Analysis of GS-ML learning rates}

In GS-ML, the training set includes all 
entries of the subgroup and a fraction of entries from the target
group.
Let $N_g$ and $N_s$ be the number of materials belonging to
a target group of interest (say, {\calligra p}$2mm$ )  and its 
subgroup (say, {\calligra p}1), respectively. 
In a learning curve such as Fig.~\ref{ml}d, the training set size, 
$N_{\rm train}$, is always greater than $N_s$.

In this Appendix, we comment on the mathematical nature of the
learning curve when $N_{\rm train} \le N_s$. In this domain, 
the training set includes part of the subgroup entries while the  
out-of-sample set contains entries from both 
the target group and the subgroup. 
Hence, the learning rate in the first ({i.e.,} $N_{\rm train} \le N_s$) half
of a complete learning curve is different than that in 
the second half. As a result, exactly at 
$N_{\rm train} = N_s$ the learning curve exhibits a discontinuity.

\begin{figure}[!hpt]
    \centering
    \includegraphics[width=0.45\textwidth]{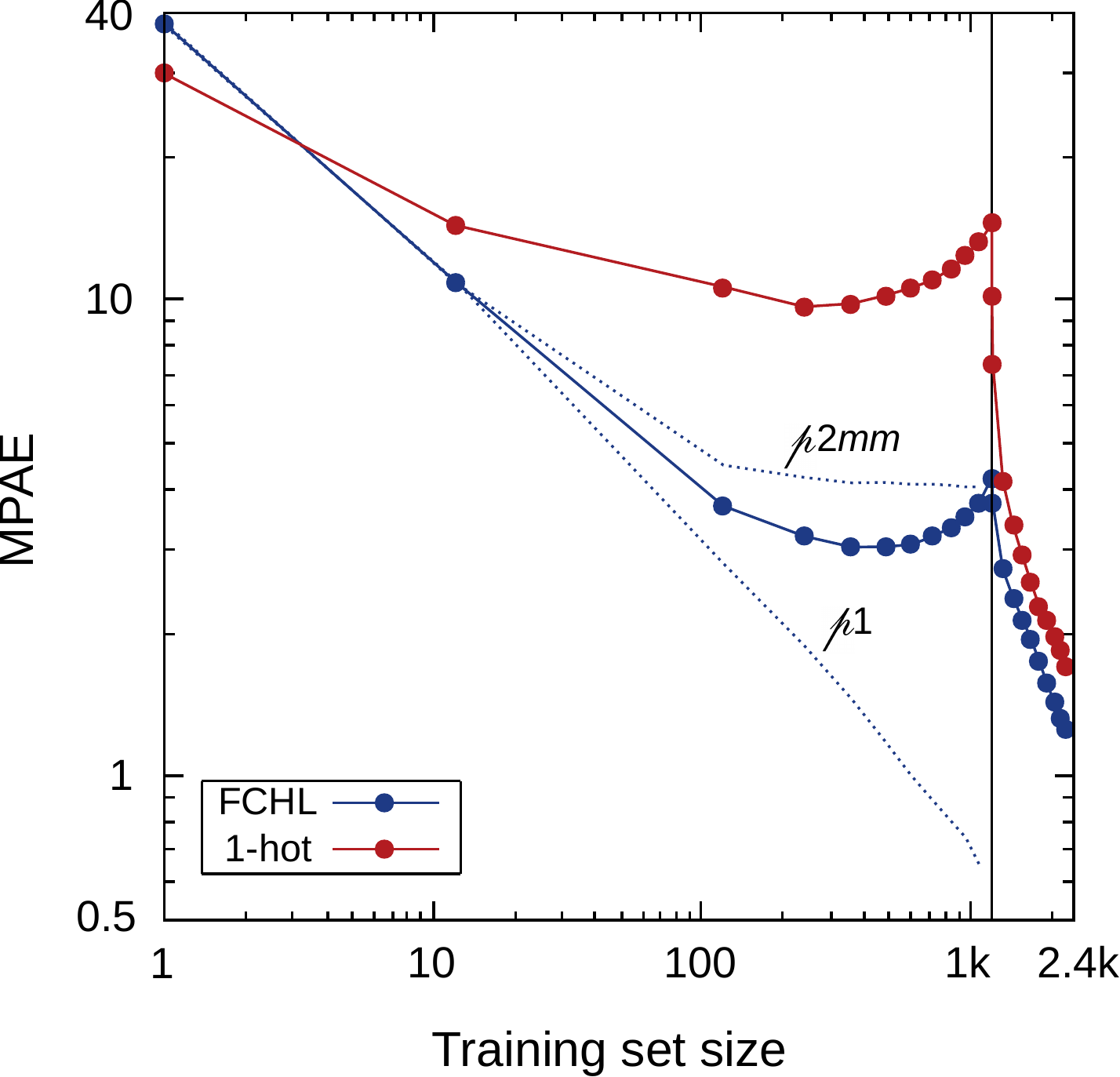}
    \caption{
    Learning curves for 1-hot (red) and FCHL (blue) based GS-ML modeling of
    FriezeRMQ1D materials of {\calligra p}1 and {\calligra p}$2mm$ symmetry. 
    Out-of-sample mean percentage absolute errors (MPAE) are shown. Individual errors for
     {\calligra p}1 and {\calligra p}$2mm$ subsets are shown in dotted lines. 
    Black vertical line marks training set size 1199.
    }
    \label{learningcurvegroups}
\end{figure}

This discontinuity is seen in Fig.~\ref{learningcurvegroups} 
at $N_{\rm train} = N_s=1199$  
in a complete learning curve irrespective of whether the representation is 
1-hot or FCHL.  Furthermore, the shape of the
graph in the first-half ($N_{\rm train} \le 1199$ ) 
of the learning curve is indicative of heteroscedasticity
in the dataset, {\it i.e.}, out-of-sample entries containing of
two classes of entries with different error distributions.
The parabolic shape of the learning curve is an outcome of
the out-of-sample mean percentage absolute error (MPAE) 
being a weighted average of individual errors for
{\calligra p}1 and {\calligra p}$2mm$:
\begin{eqnarray}
    {\rm MPAE} & = & \left[\frac{1199}{N_{\rm test}}\right]{\rm MPAE}
    ({\rm group}) + \nonumber \\ 
    & & \left[\frac{1199-N_{\rm train}}{N_{\rm test}}\right]{\rm MPAE}({\rm subgroup})
    \label{ref:eqappendix}
\end{eqnarray}
On a side note, it is of interest to note from the individual errors
projected out (dotted lines in Fig.~\ref{learningcurvegroups}) that
the {\calligra p}1 component showing a monotonic drop in the MPAE. This is because the trainingset includes examples only from {\calligra p}1.
The predictions on {\calligra p}$2mm$ is done entirely using the
structural and compositional information from {\calligra p}1, hence the {\calligra p}$2mm$ 
component of the learning curve saturates to a finite error. This finite
value corresponds to the offset of the GS-ML learning curve in Fig.~\ref{ml}d.
When $N_{\rm train} > 1199$, all entries of {\calligra p}1 are in training,  
resulting in a single out-of-sample error distribution corresponding to
materials in the {\calligra p}$2mm$ frieze group.

\bibliography{aipsamp}
\end{document}